\documentclass[reprint, superscriptaddress, amsmath, amssymb, aps, prl, twocolumn]{revtex4}
\usepackage{graphicx}

\begin{document}

\title{Effects of cell-cell adhesion on migration of multicellular clusters}

\author{Ushasi Roy}
\email{royu@purdue.edu}
\affiliation{Department of Physics and Astronomy, Purdue University, West Lafayette, IN 47907, USA}

\author{Andrew Mugler}
\email{amugler@purdue.edu}
\affiliation{Department of Physics and Astronomy, Purdue University, West Lafayette, IN 47907, USA}

\begin{abstract}
Collections of cells exhibit coherent migration during morphogenesis, cancer metastasis, and wound healing. In many cases, bigger clusters split, smaller sub-clusters collide and reassemble, and gaps continually emerge. The connections between cell-level adhesion and cluster-level dynamics, as well as the resulting consequences for cluster properties such as migration velocity, remain poorly understood. Here we investigate collective migration of one- and two-dimensional cell
clusters that collectively track chemical gradients using a mechanism based on contact
inhibition of locomotion. We develop both a minimal description based on the lattice gas model of statistical physics, and a more realistic framework based on the cellular Potts
model which captures cell shape changes and cluster rearrangement. In both cases, we find that cells have an optimal adhesion strength that maximizes cluster migration speed. The optimum negotiates a tradeoff between maintaining cell-cell contact and maintaining cluster fluidity, and we identify maximal variability in the cluster aspect ratio as a revealing signature.
Our results suggest a collective benefit for intermediate cell-cell adhesion.
\end{abstract}

\maketitle

\section{Significance Statement}
Cells have been observed to migrate faster and more efficiently in clusters than as individuals. We conjecture that adhesion among cells and with the extracellular environment plays an important role in achieving higher speed for the entire cluster. We carry out our analyses analytically and computationally, by employing a simplistic one-dimensional model and a realistic two-dimensional model which capture the essential features of multicellular migration. Our study demonstrates that an optimal cell-cell adhesion, which corresponds to maximal cellular rearrangement and loose packing, leads to a higher migration velocity for a multicellular cluster, acting as a crucial factor in effective movement of a collection of cells in a coordinated and directed fashion.

\section{Introduction}
Collective cell migration is of critical importance in nearly all stages of life \cite{Collins2015}. Biological processes like embryogenesis, morphogenesis, neurogenesis, regeneration, wound healing, and disease propagation such as cancer metastasis involve numerous cells acting in a coordinated way \cite{Friedl2009, Anona2012, Collins2015}. Studies have demonstrated that multicellular clusters can sense chemoattractants more efficiently and precisely than their isolated constituent cells do \cite{Theveneau2010, Ellison2016}.  Sensory information is combined with mechanochemical mechanisms, including actin polymerization and contact-dependent polarity (known as contact inhibition of locomotion, CIL) \cite{Theveneau2010,Mayor2010}, to produce directional migration. Recent studies have indicated that cadherin- and integrin-based adhesions at cell-cell junctions and cell-extracellular matrix (ECM) contacts respectively are indispensable for migration of multicellular clusters \cite{Collins2015, Pokutta2007, Zamir2001}. Cell-cell and cell-ECM adhesion are integrated with actin dynamics to keep clusters together during collective cell migration \cite{Collins2015, Canel2013}.

Collective migration presents a mechanical tradeoff, as cells must negotiate a balance between displacing themselves with respect to the ECM, but not separating themselves from other cells. In many cases this results in clusters that are dynamic and loosely packed rather than rigidly structured. For example, in the case of neural crest cells, a group of pluripotent cells in all vertebrate embryos that can migrate very long distances, bigger clusters split, smaller sub-clusters collide and reassemble, and gaps continually appear and disappear \cite{Richardson2016, Theveneau2010}. This raises the question of whether there is an intermediate, rather than very strong or weak, adhesion strength that optimally negotiates this tradeoff and results in dynamic loose clustering and maximally efficient collective migration. Cell adhesion is clearly crucial to collective migration, but the mechanisms are not yet well understood.

Here we use mathematical modeling and simulation to investigate the role of cell-cell and cell-ECM adhesion strength in determining collective migration efficiency and the concomitant effects on cluster shape and dynamics. Rather than focusing on the details of the mode of action or molecular properties of different types of adhesion molecules, we develop a generic model which explores the different regimes of adhesion strength, so that we may have a general understanding of the phenomena.
We start with a one-dimensional model based on the lattice gas model of statistical physics \cite{Fronczak2013} that allows us to analytically probe the collective migration velocity of a linear chain of cells as a function of adhesion strength.
We then extend this model to two dimensions using the cellular Potts model \cite{Glazier1993, Graner1992, Camley2017jpd}, which more realistically captures cell shape, cluster rearrangement, and other essential aspects of cluster migration. \

Numerical results from both the one- and the two-dimensional model suggest the existence of an intermediate adhesion strength among cells that leads to the fastest migration of a multicellular cluster. Specifically, there exists a regime of intercellular and cell-ECM adhesion strengths which corresponds to optimally effective migration. We demonstrate that, in this regime, the clusters possess the maximal rearrangement capacity while remaining as a connected cluster, rather than falling apart and scattering into single isolated cells or strongly sticking together as a compact structure.

\section{Methods}
We first consider a simplified one-dimensional model for collective migration based on the lattice gas model of statistical physics, and then a more realistic two-dimensional model based on the cellular Potts model. Here we first review the lattice gas model (later, in the Results section, we discuss our new calculations using this model, as well as our own modifications to it). We then present the model details of the cellular Potts model.

\subsection{One-dimensional lattice gas model}
\begin{figure}
\begin{center}
\includegraphics[width=\columnwidth]{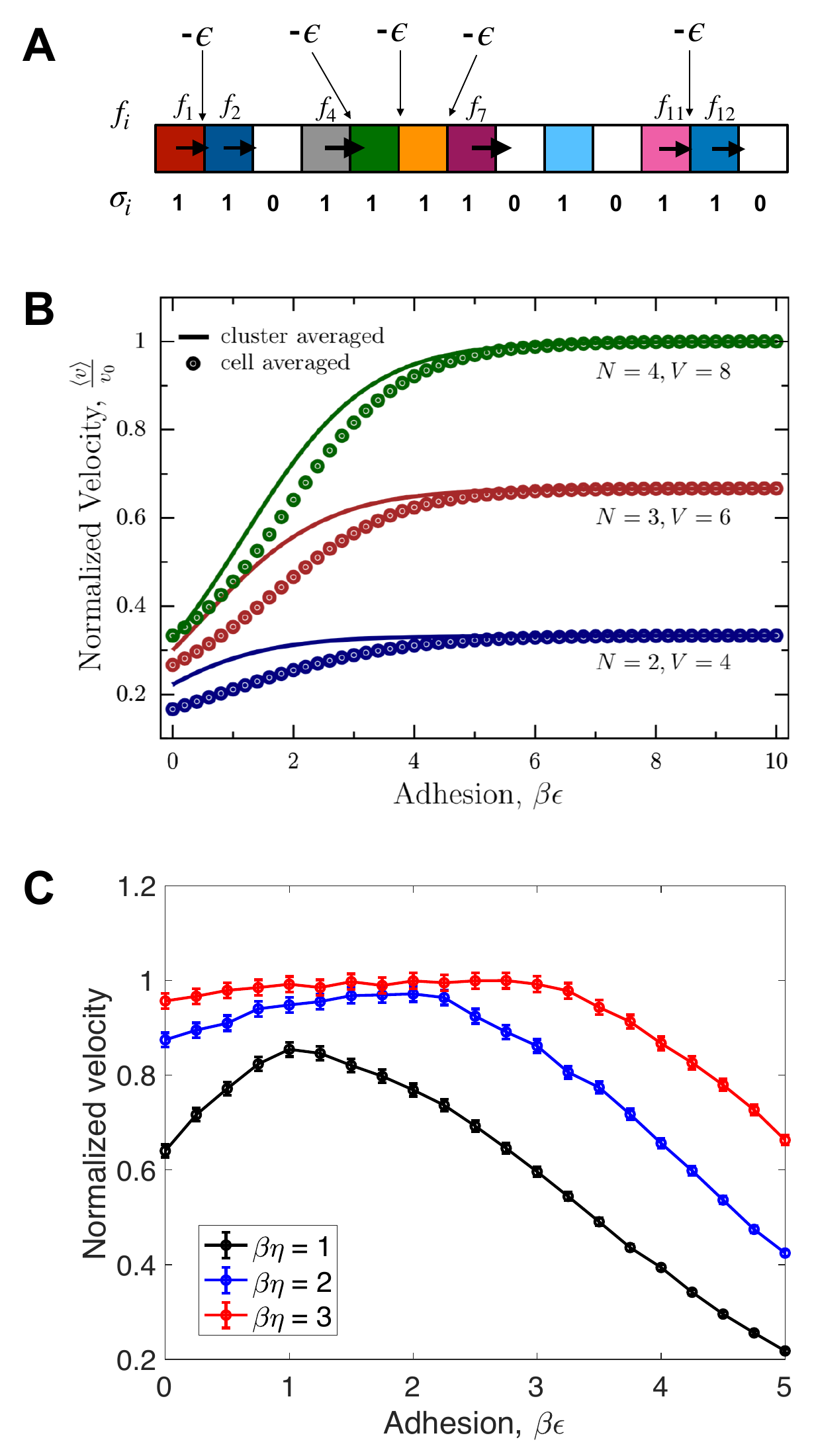}
\caption{{\bf Velocity vs.\ adhesion for one-dimensional collective cell migration.}
A. Schematic showing a collection of cells (colors, $\sigma_i = 1$) and ECM (white, $\sigma_i = 0$) arranged in a linear chain. Each pair of cells has an interaction energy $-\epsilon$. Arrows indicate motility force $f_i$. B. Normalized velocity $\langle v \rangle / v_0$ as a function of adhesion $\beta \epsilon$ for the undriven model, Eq.\ \eqref{ELG}. C. Normalized velocity as a function of adhesion $\beta \epsilon$ for the driven model, Eq.\ \eqref{Elgcpm}.}
\label{Fig1dLG}
\end{center}
\end{figure}

We first investigate a one-dimensional collective of cells using the lattice gas model. Consider $N$ cells arranged in a one-dimensional lattice of $V$ sites with $V \ge N$ (Fig.\ \ref{Fig1dLG}A). 
$\sigma_i$ denotes the state of each lattice site $i$. $\sigma_i = 1$ represents a cell while ECM is labeled by $\sigma_i=0$.

Assume that interaction exists only between adjacent cells; the total energy for a given configuration of cells $\{ \sigma_i \}$ can then be expressed as
\begin{equation} \label{ELG}
E_{LG} =-\epsilon \sum_{i=1}^V \sigma_i \sigma_{i+1}
\end{equation}
where $-\epsilon$ is the interaction energy between two adjacent cells representing their adhesion. We impose $\sigma_{V+1} = \sigma_1$ for periodicity and $\sum_{i=1}^V \sigma_{i} =  N $  to conserve cell number.

The grand partition function for the lattice gas is
\begin{equation} \label{Xilg}
\Xi_{LG} = \sum_{N=0}^V z^N Z_{LG}
\end{equation}
where $Z_{LG}=\sum_{\{\sigma_i \}} e^{- \beta E_{LG}}$ is the canonical partition function, $z = e^{\beta \mu}$ is the fugacity parameter, with $\beta=(k_B T)^{-1}$ and $\mu$  denoting the chemical potential.
The inverse of Eq.\ \eqref{Xilg} is
\begin{equation} \label{ZLGXiLG}
Z_{LG} = \frac{1}{N!} \frac{\partial^N}{\partial z^N} \Xi_{LG}.
\end{equation}
Inserting Eq.\ \eqref{ELG} into Eq.\ \eqref{Xilg} and exploiting the fact that $N = \sum_{i=1}^V \sigma_{i}$, Eq.\ \eqref{Xilg} can be recast as
\begin{equation} \label{Xilg2}
\Xi_{LG} = \sum_{\{\sigma_i \}} \exp \left( \beta \epsilon \sum_{i=1}^V \sigma_i \sigma_{i+1} +  \beta \mu \sum_{i=1}^V \sigma_i \right).
\end{equation}
We now recognize that the grand partition function of the lattice gas model as expressed in Eq.\ \eqref{Xilg2} has the same form as the canonical partition function of the Ising model  \cite{Fronczak2012, Fronczak2013}. Specifically, relating the $\sigma_i \in \{0,1\}$ to Ising spin variables $s_i \in \{-1,1\}$ via $\sigma_i = (s_i + 1)/2$, Eq.\ \eqref{Xilg2} reads
\begin{equation} \label{XiLGZI}
\Xi_{LG} = Z_I e^{\beta \mu V/2}e^{\beta \epsilon V/4},
\end{equation}
where $Z_I$ is the canonical partition function of the Ising model with magnetic field $H = (\epsilon + \mu)/2$ and coupling energy $J = \epsilon/4$.

The canonical partition function of the Ising model is exactly solvable in one dimension and reads
\begin{equation} \label{ZI}
Z_I = \lambda_+^V + \lambda_-^V
\end{equation}
for a periodic chain, where
\begin{equation} \label{lambdaI}
\lambda_{\pm} = e^{\beta J} \cosh(\beta H) \pm \sqrt{e^{2 \beta J} \sinh^2(\beta H) + e^{-2 \beta J}}.
\end{equation}
Thus, Eqs.\ \eqref{ZLGXiLG} and \eqref{XiLGZI}-\eqref{lambdaI} constitute an analytic expression for the canonical partition function of the lattice gas model.
We use this fact to calculate the cluster migration velocity in the Results section.

\subsection{Two-dimensional cellular Potts model}
To more realistically model cluster migration in two dimensions, we use computer simulation.
Many cellular automata models have been developed for this task \cite{Ermentrout1993, Maire2015, Mente2015}; we use the cellular Potts model (CPM) \cite{Szabo2010, Albert2016}. The CPM captures realistic properties such as changes in cell shape and cell size, rearrangement of cells within a cluster, and the dynamic breakup or re-aggregation of sub-clusters. Diverse biological phenomena like chemotaxis, cell sorting, endothelial cell streaming, tumor invasion and cell segregation have been modeled using the CPM \cite{Kabla2012, Szabo2010, Maclaren2015}.

\begin{figure*}
\begin{center}
\includegraphics[width=\textwidth]{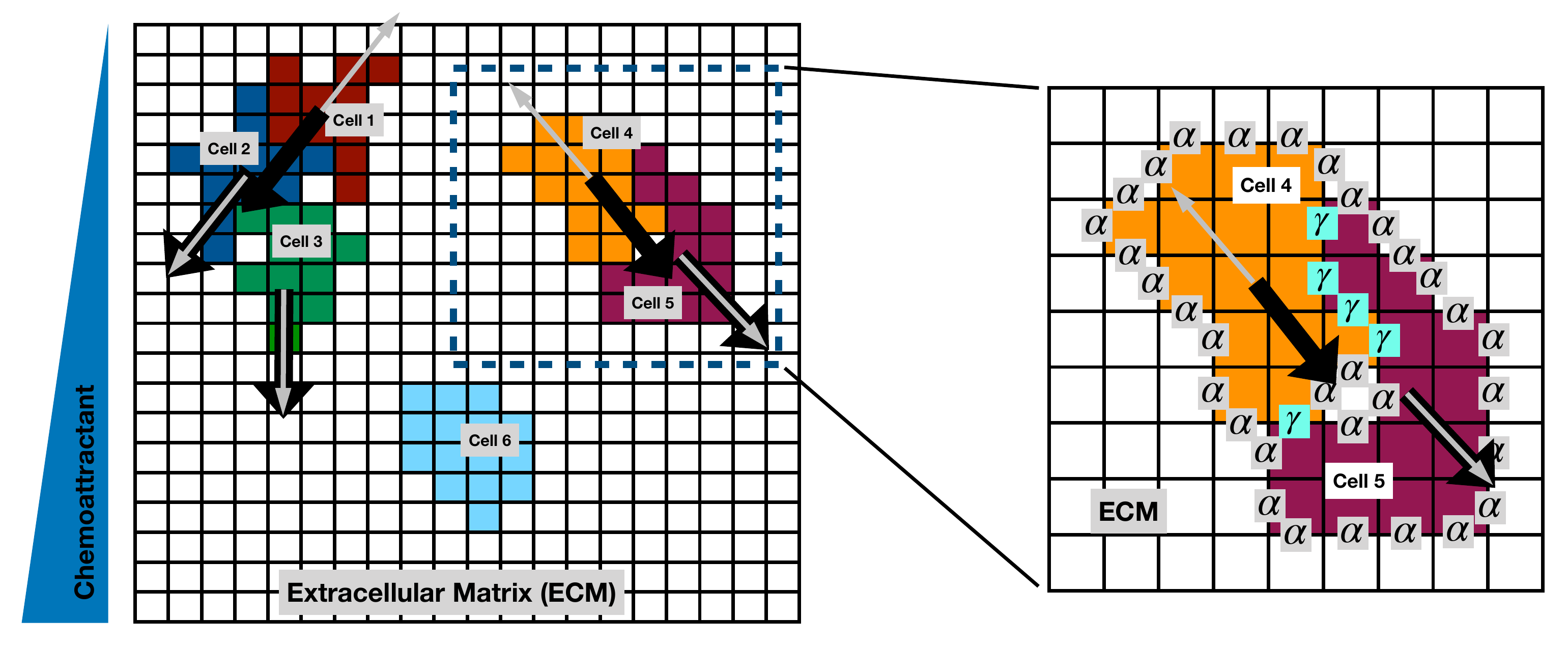}
\caption{{\bf Cellular Potts model for collective migration in a chemical gradient.} A schematic of the adaptive cellular Potts model (CPM) depicting a characteristic snapshot of three multicellular clusters of different sizes. The cluster consisting of two cells, enclosed within a dashed box (left), is zoomed (right) to show cell-cell energy penalty $\gamma$ and cell-ECM energy penalty $\alpha$. All cells have respective motility force vectors (black arrows) and repulsion vectors (gray arrows; away from cell-cell contact as a result of CIL) in a linear chemoattractant gradient. A single isolated cell (cell 6) has no force acting on it since we have considered CIL as our guiding mechanism for motility.}
\label{CPMschematic}
\end{center}
\end{figure*}

We have considered a discrete two-dimensional lattice. Each cell is represented by a group of lattice sites $x$ with the same integral values for their lattice labels $\sigma(x)>0$ (Fig.\ \ref{CPMschematic}). The empty lattice sites correspond to the extra-cellular matrix (ECM), with lattice label $\sigma(x)=0$, providing an environment through which the cells move. The initial configuration has several cells arranged in a single cluster. The energy of the whole system $E_{CPM}$ has contributions from two factors: the first one is the adhesion while the second one is the area restriction term,
\begin{equation} \label{ECPM}
E_{CPM} = \sum_{\langle x,x' \rangle} J_{\sigma(x),\sigma(x')} + \sum_{i=1}^N \lambda_A(\delta A_i)^2.
\end{equation}
The adhesion energy term $J_{\sigma(x),\sigma(x')}$ is given by the following
\begin{equation}
\small{
J_{\sigma(x),\sigma(x')} =
\begin{cases}
 0 \;\;\;\;\;\, \sigma(x)\sigma(x) \; \ge 0 & \; \textrm{within ECM or same cell},  \\
\alpha \;\;\;\;\;  \sigma(x)\sigma(x') = 0 &\;  \textrm{cell-ECM contact}, \\
  \gamma \;\;\;\;\; \sigma(x)\sigma(x') > 0 & \; \textrm{cell-cell contact}.
  \end{cases}
  }
\end{equation}
$\alpha$ denotes the interaction strength of any cell due to adhesion with its environment while intercellular adhesiveness is characterized by $\gamma$.
A migrating cell is refrained from growing or shrinking to unphysical sizes, as well as branching or stretching into unphysical shapes, due to the presence of the area restriction term in Eq. \eqref{ECPM}. Cells undergo fluctuations in size $\delta A_i$ around a desired area $A_0$ via $\delta A_i \equiv A_i(t) - A_0$. We have set $\lambda_A$ to be unity \cite{Varennes2016}.
Previous work \cite{Glazier1993, Graner1992, Varennes2016, Camley2017jpd} has included a perimeter restriction term in addition to the area restriction term. For simplicity we omit this term, as we find that sufficiently large $\alpha$ and $\gamma$ constrain perimeter by cell-ECM or cell-cell contact.

Our model of migration is based on contact inhibition of locomotion (CIL), a well known and central mechanism of collective cell movement \cite{Mayor2010}. The formation of cell protrusions is locally inhibited when a cell comes into contact with another cell, and hence the cell ceases to move in that direction. Instead, the cell generates protrusions away from the site of contact \cite{Fontaine2008, Desai2013}, which produces force in the outward direction. Direct evidence of CIL has been observed in migrating clusters, where outer cells have strong outward polarization while inner cells weakly protrude \cite{Theveneau2010}. Note that under this mechanism, directional migration is purely collective: two or more cells in contact are polarized, whereas single isolated cells are not.

We consider the case where cells exist in an external chemical gradient. {\it Drosophila} egg chamber cells \cite{Bianco2007, Rorth2007, Inaki2012, Wang2010}, clusters of lymphocytes \cite{Engra2015}, neural crest cells \cite{Theveneau2010}, and epithelial organoids \cite{Ellison2016} exhibit emergent gradient sensing and collective migration in response to graded chemical cues.
Under the assumption that the chemical concentration influences the magnitude of the protrusive forces, the presence of a chemical gradient creates a force imbalance \cite{Camley2016a, Camley2016b}, allowing the cluster to respond to the gradient. However, as a cluster migrates up a gradient according to this mechanism, the background concentration increases, which increases the outward forces and can cause the cluster to scatter \cite{Camley2016a}. To prevent scattering, we adopt an adaptive mechanism of gradient sensing \cite{Camley2016a, Ellison2016, Varennes2016}, in which cells respond to the difference between the local chemical concentration and the average experienced over the entire cluster. Evidence for adaptive collective gradient sensing has been observed in epithelial organoids \cite{Ellison2016}.

Specifically, we take the magnitude of the force experienced by cell $i$ to be
\begin{equation} \label{FCPM}
F_i = \eta g (x_{cm} ^i - x_{ccm})
\end{equation}
where $\eta$ sets the force strength, $g$ is the concentration gradient which is in the $x$ direction (downward in Fig.\ \ref{CPMschematic} and subsequent figures), $x_{cm}^i $ and $x_{ccm}$ are the $x$ coordinates of the center-of-mass of the cell and of the whole cluster respectively, and the subtraction expresses the adaptivity.
We set $\eta g = 1$ in this work.
The direction of the force experienced by cell $i$ is determined according to CIL \cite{Varennes2016}: we sum all vectors pointing from cell-pixels in contact with any other cell to the center-of-mass of cell $i$. This net `repulsion' vector points outward (gray in Fig.\ \ref{CPMschematic}), whereas the force direction is flipped when the sign of Eq.\ \eqref{FCPM} is negative (black in Fig.\ \ref{CPMschematic}).
The forces contribute a work term to the energy functional, given by
\begin{equation} \label{biasW}
W = - \sum_{i=1}^N \vec{F}_i \cdot \Delta\vec{x}_i,
\end{equation}
where $\Delta\vec{x}_i$ is the change in the center-of-mass of each cell upon a configurational change, discussed next.

Given the energy and work terms, cellular dynamics under the CPM are simulated using a Monte Carlo process which is based on the principle of minimizing the energy of the whole system. Specifically, motility is modeled by an addition (copying the identity of one cell-pixel, chosen randomly, to its neighboring site) or removal (copying an ECM-pixel to a site previously occupied by cell) of pixels. Each Monte Carlo step selects randomly a pair of adjacent lattice sites, and attempts to copy the identity of one to the other. It calculates the energy of the previous (before copying) and the new (after copying) configuration. The new configuration is accepted with probability $P$, given by
\begin{equation}
P = 
\begin{cases}
e^{-(\Delta E_{CPM} + W) } \;\;\; & \Delta E_{CPM} + W \ge 0 \\
1 \;\;\; & \Delta E_{CPM} + W < 0,
\end{cases}
\end{equation}
where $\Delta E_{CPM}$ is the change in energy of the system due to the attempted move, calculated from Eq. \eqref{ECPM}, and $W$ is the bias term given by Eq. \eqref{biasW}.

\section{Results}

\subsection{Driven lattice gas model exhibits optimal cell-cell adhesion}

We first consider the one-dimensional lattice gas model (Methods) and ask how the average cell velocity depends on the adhesion strength. As in the CPM described above, we assume that the force ($f_i$ in Fig.\ \ref{Fig1dLG}A) is exerted by the edge cells due to CIL and is proportional to the local concentration of an external chemical. In one dimension, there are only two edge cells per sub-cluster of at least two cells (single isolated cells experience no contacts and therefore no force). In a linear chemical profile, the net force will be proportional to the linear extent of the sub-cluster, equivalent to the number of cell-cell contacts. Assuming that the velocity is proportional to the force (appropriate at low Reynolds number), the velocity of a sub-cluster can be expressed as $v_0 \sum_i\sigma_i\sigma_{i+1}$, where the sum extends over the indices of the sub-cluster, and $v_0$ is an arbitrary constant that sets the velocity scale. The average velocity over all sub-clusters in a particular configuration $\{\sigma_i\}$ is the sum of all such terms divided by the total number of sub-clusters, or
\begin{equation} \label{vconfig}
v = \frac{v_0\sum_{i=1}^V \sigma_i \sigma_{i+1} }{\sum_{i=1}^V  \sigma_i (1- \sigma_{i+1})} = -\frac{v_0 E_{LG}} { \epsilon N + E_{LG} }.
\end{equation}
Here the denominator counts sub-clusters by their rightmost edges, and the second step recalls Eq.\ \eqref{ELG}. We have chosen to weight each cluster equally in Eq.\ \eqref{vconfig} for analytic tractability, but we will see that similar results are obtained if each cell is weighted equally instead, as in later Results sections.

The average velocity is the sum of Eq.\ \eqref{vconfig} against the Boltzmann probability,
\begin{equation} \label{AvgVel}
\langle v \rangle
=  \sum_{\{\sigma_i\}} \frac{-v_0 E_{LG}}{\epsilon N + E_{LG} } \times \frac{e^{- \beta E_{LG}}}{Z_{LG}}
= \frac{v_0}{Z_{LG}} \sum_{n=0}^{\infty} \left( \frac{\partial_{\beta}}{\epsilon N}  \right)^n Z_{LG}.
\end{equation}
The second step recognizes that $n$ derivatives of the partition function extract $n$ powers of $-E_{LG}$, which when summed as a geometric series are equivalent to the first expression.
Eq.\ \eqref{AvgVel} connects the average velocity with the canonical partition function of the lattice gas, for which we have an analytic expression (Methods).

Eq.\ \eqref{AvgVel} depends on the size of the lattice $V$, the number of cells $N$, the velocity scale $v_0$, and the dimensionless adhesion energy $\beta\epsilon$. Therefore, we can ask for a given $V$ and $N$, how the normalized velocity $\langle v\rangle/v_0$ depends on the adhesion strength $\beta\epsilon$.
As an example,
for $V = 8$ and $N = 4$, Eq.\ \eqref{AvgVel} evaluates to
\begin{equation}
\frac{\langle v \rangle}{v_0} 
= \frac{4 e^{\beta \epsilon} + 18 e^{2\beta \epsilon} + 12 e^{3\beta \epsilon} }{1 + 12 e^{\beta \epsilon} + 18 e^{2\beta \epsilon} + 4 e^{3\beta \epsilon} }.
\end{equation}
We see in Fig.\ \ref{Fig1dLG}B (green curve) that $\langle v\rangle/v_0$ is a monotonically increasing function of $\beta\epsilon$. 

In general we find analytically that velocity increases monotonically with adhesion strength for other values of $N$ and $V$, and also numerically when cells are weighted equally in the average (Fig. \ref{Fig1dLG}B). This would imply that the optimal adhesion is infinitely strong. 
However, thus far,
this model neglects the impact of the motility process itself on the probability of occurrence of each configuration $\{\sigma_i\}$. That is, the probability is determined entirely by the Boltzmann distribution, which depends only on the adhesion energy. Instead, we expect that the motility forces will influence the ensemble of configurations, as some configurations that are driven by collective movement will occur more frequently than they would in the undriven system.

To account for the influence of motility on the configuration ensemble, we add a driving term to the energy function that is proportional to the motility forces. Specifically, we consider the change in energy to be of the following form,
\begin{equation} \label{Elgcpm}
\Delta E = \Delta E_{LG}  - \eta f_i \Delta x.
\end{equation}
Here $\Delta E$ is the change in energy when cell $i$ shifts to a neighboring lattice position. $\Delta E_{LG}$ is the change in the adhesion energy according to Eq.\ \eqref{ELG}, while $- \eta f_i \Delta x$ is the work that occurs when the change in cell position $\Delta x$ aligns with the motility force $f_i$. The latter term is analogous to the work term in the CPM, Eq.\ \eqref{biasW}. The sign of this term reflects the fact that the motility forces on both ends of the cluster point in the gradient direction, due to the adaptivity (see Methods for details). We continue to take $f_i=n-1$ to be the number of connected edges in the sub-cluster of size $n$, and $\eta$ sets the strength of the motility. Note that $\eta = 0$ corresponds to the undriven ensemble as before.

We evolve the system via Monte Carlo simulation as in the CPM (Methods). We randomly choose a pair of non-identical neighboring sites, i.e., a cell and an ECM site, and swap them, calculate the energy change following Eq.\ \eqref{Elgcpm}, and accept the new configuration with Boltzmann probability $e^{-\beta \Delta E}$. The center-of-mass velocity averaged over many instances is shown in Fig.\ \ref{Fig1dLG}C for different values of $\beta\eta$. We observe in all cases that there is a clear optimum in the adhesion strength for which the cluster has the maximum migration velocity. We conclude that the effect of motility is to bias the ensemble of configurations away from its equilibrium distribution, which is necessary to observe an optimal adhesion strength.

The optimal adhesion strength arises due to the following tradeoff. On the one hand, weak adhesion results in isolated cells that diffuse without bias, except when they happen to collide and briefly attain a bias due to the CIL. On the other hand, strong adhesion causes the first term in Eq.\ \eqref{Elgcpm} to dominate over the second, suppressing movement of cells at the leading edges of sub-clusters, and therefore suppressing movement as a whole. The optimal adhesion strength negotiates the balance between the two, resulting in clusters that are tight enough to cohere but fluid enough to allow forward progress.

The one-dimensional model considered thus far captures the core physics of an optimal adhesion strength but necessarily neglects changes in cell and cluster shape, as well as intra-cluster cell rearrangements, that are typical of multicellular migration in larger dimensions. Therefore, we use the two-dimensional CPM to investigate these aspects next.

\subsection{Cellular Potts model exhibits optimal cell-cell and cell-ECM adhesion}
\begin{figure}
\begin{center}
\includegraphics[width=\columnwidth]{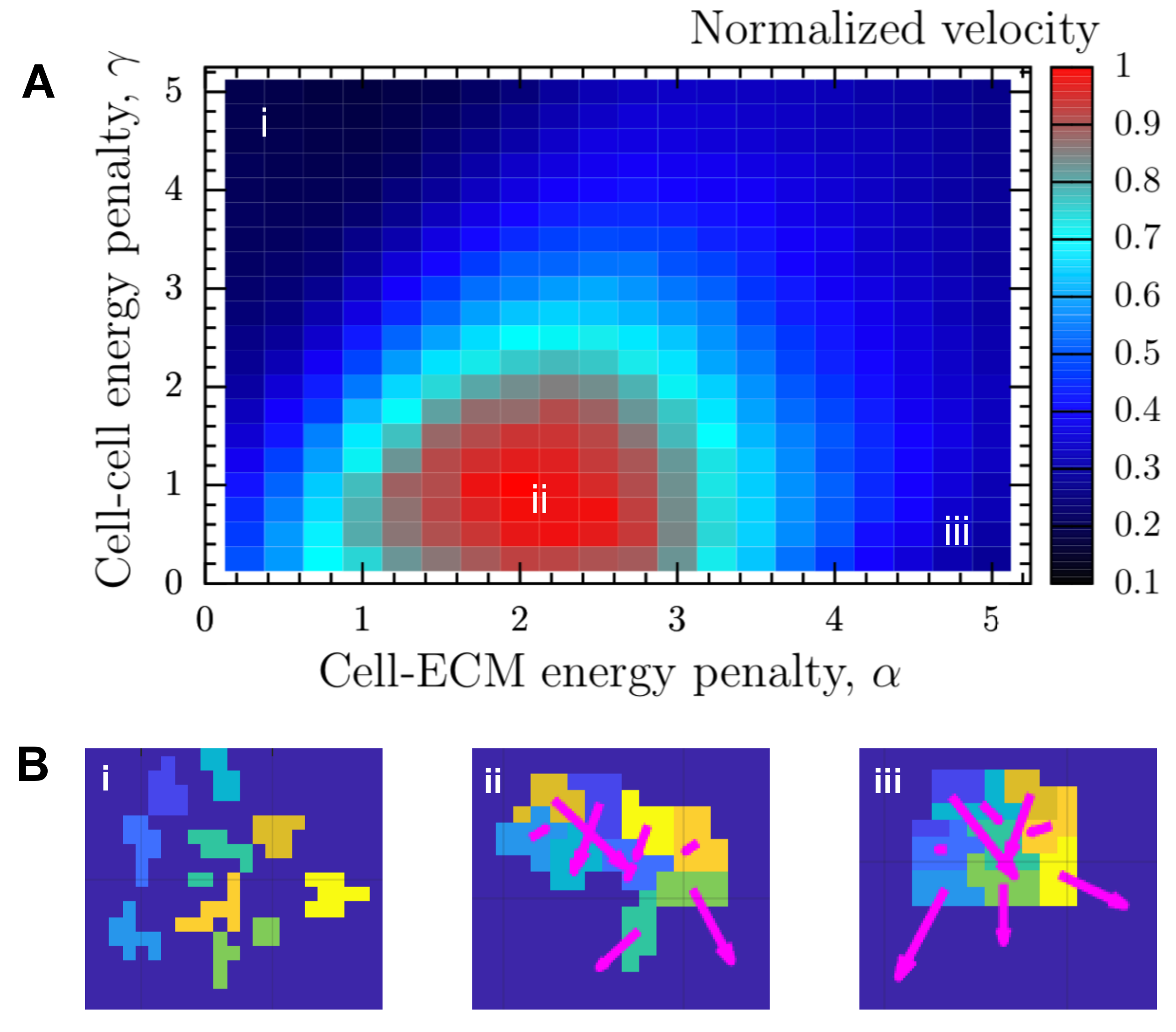}
\caption{{\bf Velocity vs.\ intercellular and cell-ECM adhesion strengths for two-dimensional collective cell migration.} A. Normalized center-of-mass velocity vs.\ cell-ECM energy penalty $\alpha$ and cell-cell energy penalty $\gamma$. Velocity is maximal in region ii. Velocity is computed after 20,000 Monte Carlo steps and averaged over 200 trials for each value of $\alpha$ and $\gamma$. B. Snapshots from simulation of a cluster of nine cells, illustrating the cluster configuration while migrating, corresponding to different regimes in the parameter space: (i) cells scatter and diffuse away, (ii) cells remain connected with intermediate adhesion, and (iii) cells tightly adhere to one another forming a compact structure.}
\label{Fig_heatmap_dist}
\end{center}
\end{figure}

To capture more realistic motion of cells in two dimensions, we use the CPM (Methods).
We plot the migration velocity for a cluster of nine cells in the phase space of $\alpha$, which represents the energy penalty for cell-ECM contact, and $\gamma$, which represents the energy penalty for cell-cell contact (see Fig.\ \ref{Fig_heatmap_dist}A). We see a clear optimum in regime ii (red), corresponding to intermediate $\alpha$ and $\gamma$. We have checked that the existence and location of the optimum is not strongly dependent on the number of cells in the system.
Thus, not only is there an optimal cell-cell adhesion strength ($\gamma$) as found in the one-dimensional model, there is also an optimal cell-ECM adhesion strength ($\alpha$).

The reason for the optimum is illustrated in Fig.\ \ref{Fig_heatmap_dist}B. At low $\alpha$ and high $\gamma$ (region i), cells adhere to the ECM but not each other. Therefore, they scatter and do not benefit from the collective determination of the gradient direction, resulting in a low velocity. At high $\alpha$ and low $\gamma$ (region iii), cells adhere to each other but prefer to avoid contact with the ECM. The latter prevents protrusions from forming, also resulting in a low velocity. Region ii optimally negotiates this tradeoff.

Although Fig.\ \ref{Fig_heatmap_dist} demonstrates the existence of optimal adhesion strengths, it does not directly address the question of what properties of the clusters correspond to this optimum. As these properties could lead to experimental predictions and further reveal the physical mechanisms behind optimal collective migration, we explore this question next.

\subsection{Optimum arises from intact but maximally fluid clusters}

\begin{figure}
\begin{center}
\includegraphics[width=\columnwidth]{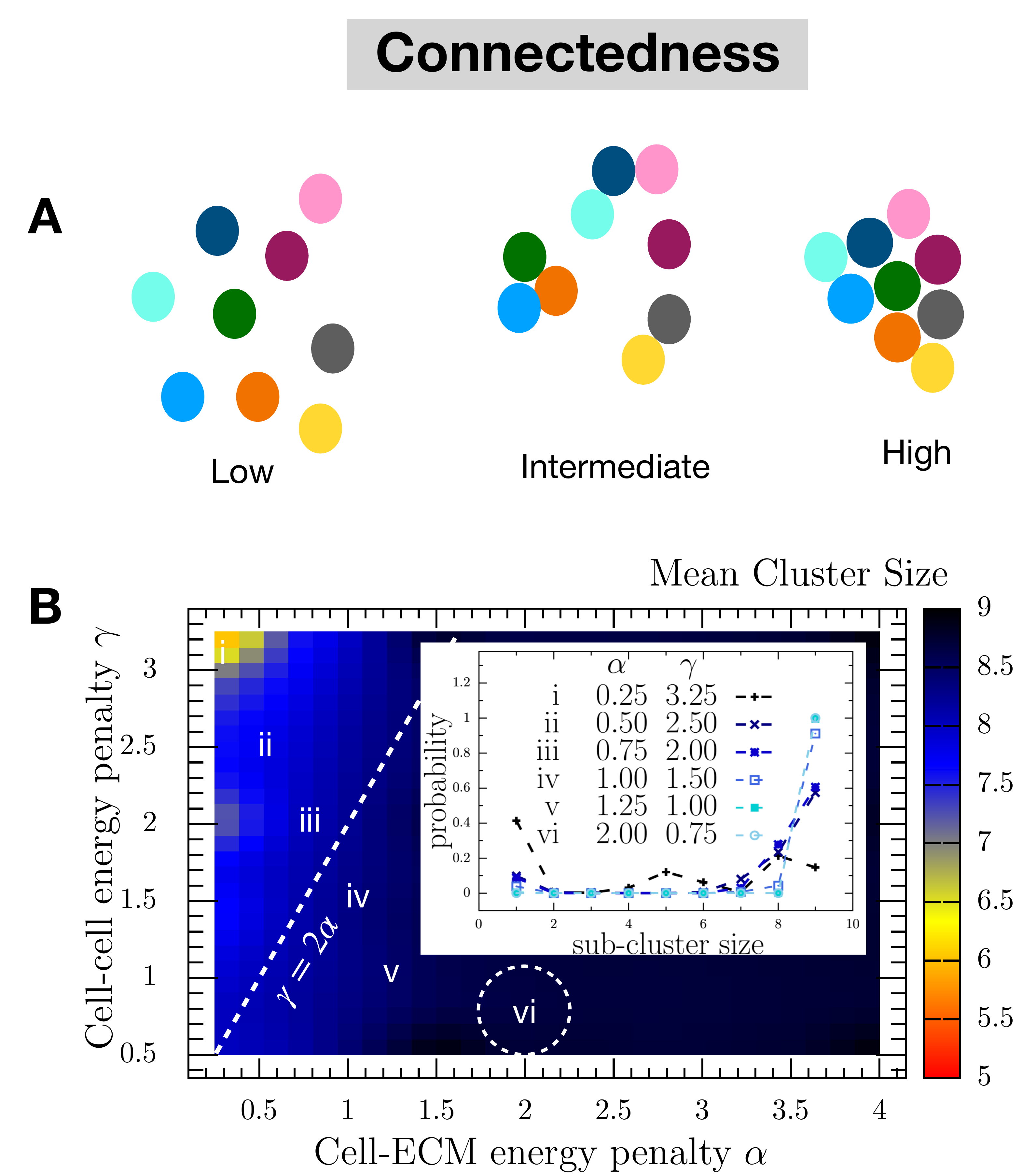}
\caption{{\bf Connectedness transition does not account for maximal cluster velocity.} A. Schematic illustrating low, intermediate, and high connectedness. B. Mean cluster size vs.\ $\alpha$ and $\gamma$ for 9 cells. Cells transition from disconnected to connected when $\alpha > 2\gamma$, as predicted, which is far from where velocity is maximal (dashed circle). Inset: Sub-cluster size distribution for different values of $\alpha$ and $\gamma$ (as shown by i-vi in A) clearly exhibits a transition from multiple sub-clusters to a single cluster of size nine. Sub-cluster sizes are computed over 10,000 Monte Carlo steps for each value of $\alpha$ and $\gamma$.}
\label{Fig4_connectedness}
\end{center}
\end{figure}

We first hypothesized that the optimal migration velocity corresponds to the transition between a fully connected cluster and multiple disconnected sub-clusters (Fig.\ \ref{Fig4_connectedness}A). Such a transition occurs when $\gamma \approx 2\alpha$. The reason is that
two cell edges that are in contact with each other will have an energy cost of $\gamma$, whereas if these two edges are exposed to the ECM they will have an energy cost of $2\alpha$. Thus $\gamma<2\alpha$ will promote cell scattering, while $\gamma>2\alpha$ will promote cluster cohesion.

Fig.\ \ref{Fig4_connectedness}B confirms the transition: we see in Fig.\ \ref{Fig4_connectedness}B that to the left of the line $\gamma = 2\alpha$ (dashed) the mean sub-cluster size is less than the total cell number of 9 cells, whereas to the right of the line it converges to 9 cells. Indeed, in the inset of Fig.\ \ref{Fig4_connectedness}B we see that far to the left of the transition (region i), the sub-cluster size distribution is broad, with significant probability to observe sub-clusters of size less than nine, including isolated cells of size one. In contrast, far to the right of the transition (region vi), we see that the sub-cluster size distribution has support only at nine, meaning all cells remain intact throughout the migration.

The optimal velocity occurs in region ii of Fig.\ \ref{Fig_heatmap_dist}A which corresponds to region vi of Fig.\ \ref{Fig4_connectedness}B (dashed circle), which is far from the connectedness transition. Evidently, being relatively deep within the fully connected regime is optimal for maximal cluster velocity. Therefore, being at the transition between connected and disconnected cannot explain the optimum observed in our model.

\begin{figure}
\begin{center}
\includegraphics[width=\columnwidth]{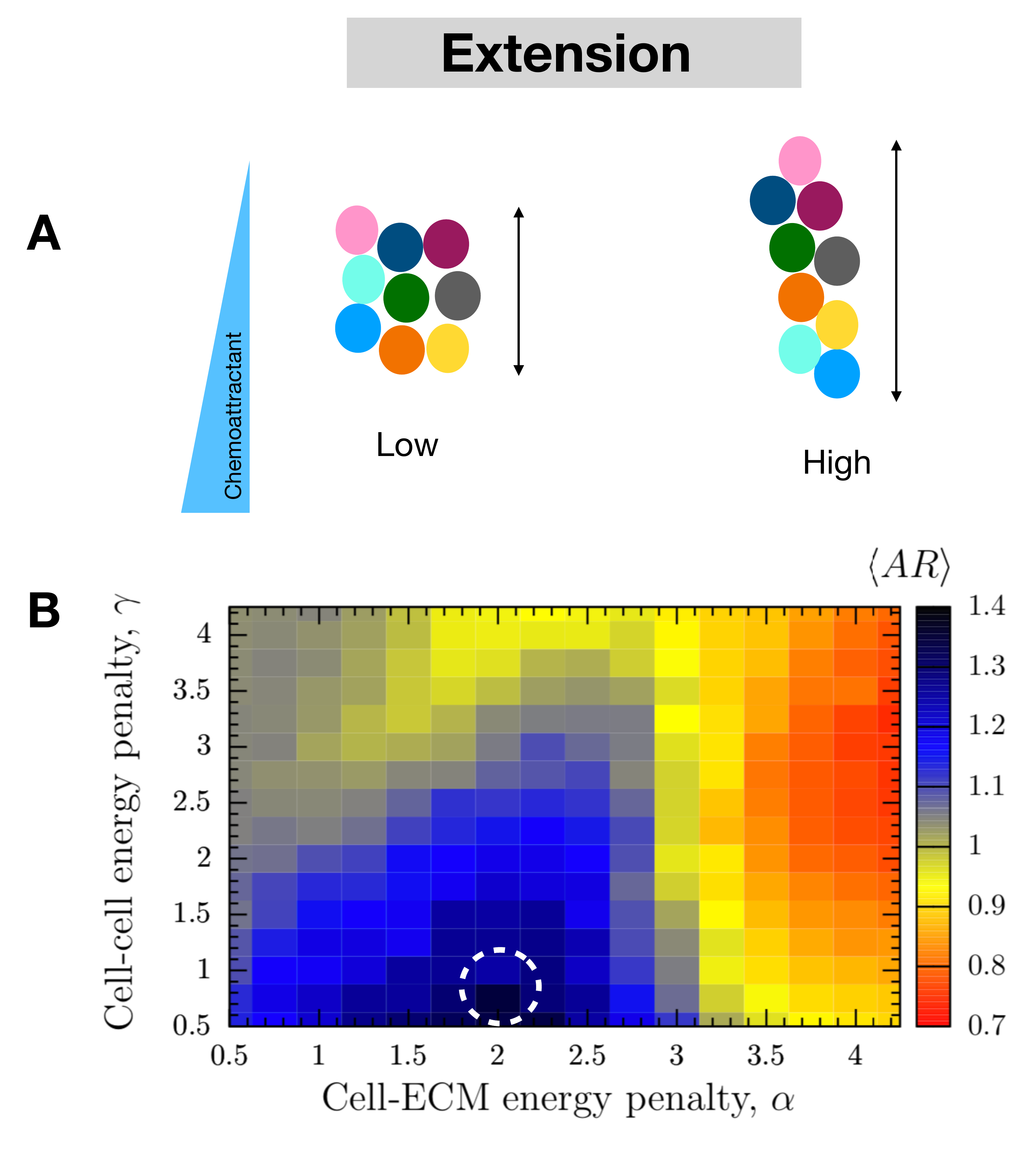}
\caption{{\bf Extension in gradient direction correlates with cluster velocity.} A. Schematic illustrating low and high cluster extension in the gradient direction, which we quantify by the aspect ratio (AR). B. Mean aspect ratio $\langle AR \rangle$ vs.\ $\alpha$ and $\gamma$ exhibits maximum in same location as maximal cluster velocity (dashed circle). Aspect ratio is computed over 20,000 Monte Carlo steps and averaged over 200 trials for each value of $\alpha$ and $\gamma$.}
\label{Fig5_extension}
\end{center}
\end{figure}

We next hypothesized that the optimal migration velocity corresponds to the ability of the cluster to  extend maximally in the gradient direction  while remaining intact (Fig.\ \ref{Fig5_extension}A). Maximal extension would allow the cluster to span the largest distance in the gradient direction, meaning that the concentration difference between the front (or back) cell and the cluster center-of-mass would be largest. This would result in the largest force exerted by these cells via Eq.\ \eqref{FCPM}. We quantify extension using the cluster aspect ratio (AR): the ratio of the length of the cluster parallel vs.\ perpendicular to the gradient direction. We see in Fig.\ \ref{Fig5_extension}B that the average aspect ratio indeed varies as a function of the adhesion parameters $\alpha$ and $\gamma$, and that a maximum is observed (dark blue) corresponding to extension parallel to the gradient direction ($\langle AR\rangle > 1$). The location of this maximum corresponds to that of the maximal velocity (dashed circle in Fig.\ \ref{Fig5_extension}B). We conclude that maximal cluster extension leads to maximal migration velocity.

\begin{figure}
\begin{center}
\includegraphics[width=\columnwidth]{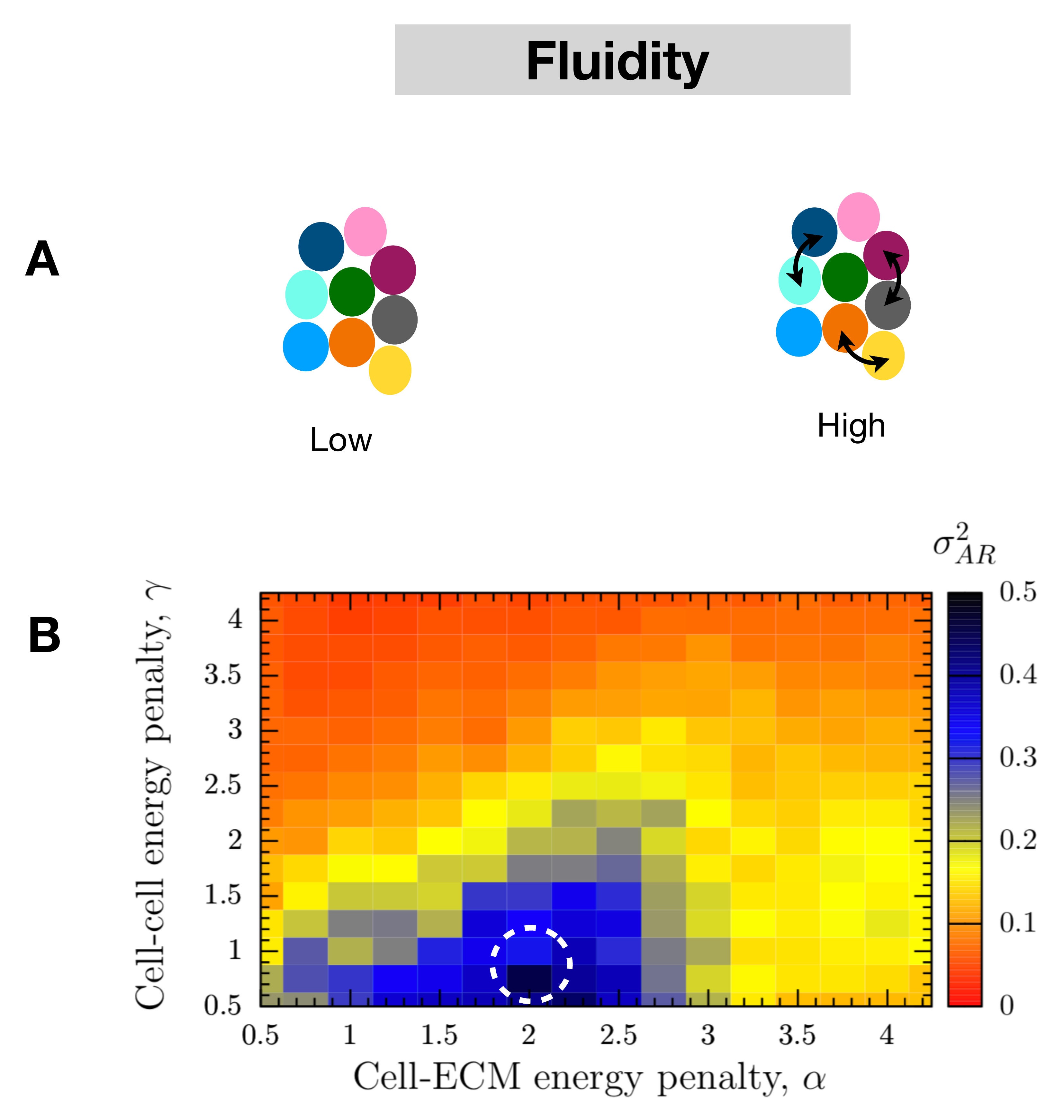}
\caption{{\bf Cluster fluidity correlates with cluster velocity.} A. Schematic illustrating low and high fluidity. High fluidity corresponds to cell rearrangement and changes in overall cluster shape, which we quantify using the variance of the aspect ratio. B. Variance of the aspect ratio vs.\ $\alpha$ and $\gamma$ exhibits maximum in same location as maximal cluster velocity (dashed circle). Aspect ratio is computed over 20,000 Monte Carlo steps, and variance is computed over 200 trials for each value of $\alpha$ and $\gamma$.}
\label{Fig6_fluidity}
\end{center}
\end{figure}

The maximal average extension observed in Fig.\ \ref{Fig5_extension}B could occur in multiple different ways. One possibility is that the cluster relaxes to a maximally extended shape and stays in this shape throughout the course of the migration. An alternative possibility is that the cluster is fluid, with cells free to rearrange while the cluster remains intact (Fig.\ \ref{Fig6_fluidity}A). Previous studies have shown that fluidity determines the properties of a jamming transition in confluent sheets \cite{Dapeng2016}, and that more fluid multicellular clusters can be more effective gradient sensors \cite{Camley2017pnas}.  If the cluster is fluid, motility forces would then drive the cluster into a maximally extended shape on average, but many shapes could be visited throughout the migration process. We therefore expect the two possibilities of a rigid or a fluid cluster to have low or high variability in the aspect ratio, respectively.

To distinguish between these two possibilities, we define fluidity using the variance in the aspect ratio, $\sigma^2_{AR}$.
Fig.\ \ref{Fig6_fluidity}B plots $\sigma^2_{AR}$ as a function of $\alpha$ and $\gamma$.
We see that it has a maximum at the same location of the optima in the migration velocity and the cluster extension (dashed circle). Thus, maximal velocity corresponds not to a cluster that is rigidly extended in the gradient direction, but to a cluster that is maximally fluid: extended on average, but freely exploring the space of cluster shapes as migration proceeds. This maximal fluidity is enabled at intermediate adhesion strengths: sufficiently strong to keep cells intact as a fully connected cluster, but sufficiently weak to allow maximal variability in cluster shape.

\section{Discussion}

We have developed a model to investigate the role of cell-cell and cell-ECM adhesion in determining the migration velocity of multicellular clusters. In our model, migration is (i) collective, based on contact inhibition of locomotion, and (ii) directed, due to the presence of an external gradient. In its simplest form---point-like cells in one dimension---we have mapped the model to the lattice gas model of statistical physics, which affords analytic results for the migration velocity. We have seen that an optimal cell-cell adhesion strength emerges that maximizes migration velocity, and that this optimum depends on the interplay between the motility forces and the configurational statistics of the cells. In its more realistic form---spatially extended cells embedded in ECM in two dimensions---we have seen that the optimum exists for both cell-cell and cell-ECM adhesion strengths. Clusters with intermediate adhesion are fastest because they are the most fluid: they are intact, extended in the gradient direction, and maximally variable in cluster shape.

Our prediction that there exist optimal cell-cell and cell-ECM adhesion strengths could be tested experimentally. Experiments suggest that both cell-cell and cell-ECM adhesion are crucial for tumor invasion, as well as for homeostasis in healthy tissues \cite{Janiszewska2020}. Experimental perturbations could be used to modulate cadherin or integrin levels to tune cell-cell or cell-ECM adhesion respectively, and the effects on migration velocity could be investigated. For example, downregulation of E-cadherin within a tumor spheroid was recently achieved by introduction of interstitial flow, which was subsequently seen to promote tumor invasion \cite{Huang2020a}.

Our observation that variability in aspect ratio correlates with migration velocity could also serve as a phenomenological signature to look for in experiments. Variability in cluster shape is straightforward to extract from microscopy videos and quantify, and it abstracts away the underlying molecular details of the adhesion or migration. It would be interesting to see whether the fastest clusters are generically the most fluid across biological systems, regardless of the nature of the molecular perturbation applied.

We have considered only one- and two-dimensional migration, whereas three-dimensional migration is clearly prevalent, rich in its modalities (e.g., mesenchymal, amoeboid, lobopodial), and dependent on tunable factors (e.g., adhesion, cell confinement, contractility, deformability, proteolytic capacity) \cite{Liu2015, Callan2016, Yamada2019}. It would be possible in the future to extend our model to three dimensions to investigate some of these factors and migration modes. Nonetheless, important examples of 1D and 2D migration exist, to which our findings may more directly apply.
Examples of 1D or quasi-1D migration include preferential migration of tumor cells, cancer stem cells, and leukocytes along a bundle of linear collagen fibrils \cite{Provenzano2006, Boissonnas2007}, as well as migration of fibroblasts on 1D fibril-like lines \cite{Doyle2009, Yamada2019}.
Examples of 2D or quasi-2D migration include wound healing (or gap closure) in an epithelial tissue, cells migrating on a bone, migration of single epithelial cells along 2D sheets of basement membranes, and patrolling of leukocytes along the luminal surface of blood vessels \cite{Weigelin2012, Fisher2018, Hsu2013, Auffray2007}.

Our observation that cluster fluidity maximizes migration velocity is a purely mechanical effect: intermediate adhesion promotes cluster configurations that maximize net motility forces in the gradient direction. Previous work has also shown that cluster fluidity improves gradient sensing due to a different mechanism: fluidity averages out detection noise due to cell-to-cell variability \cite{Camley2017pnas}. We do not consider detection noise \cite{Camley2017pnas, Varennes2016, Ellison2016} or cell-to-cell variability \cite{Camley2017pnas} here. It would be interesting to investigate how these distinct advantages of cluster fluidity act in concert or whether they combine synergistically.

The model developed here is generic, minimal, and not specific to any particular cell type.
In general, there can be more than one cell type within a single cluster. In that case, it is straightforward to extend our model to include a set of cell-cell interaction parameters $\gamma_{ij}$ between every pair of cell types $i$ and $j$, or a set of cell-ECM interaction parameters $\alpha_k$ for each cell type.
We have considered only the simplest version of this scenario here, but it may be interesting in the future to generalize our work to systems that exhibit heterogeneous collective migration.

\section*{Acknowledgments}
UR thanks Michael Vennettilli for valuable inputs in the analytical treatment. UR acknowledges helpful discussions with Raj Kumar Manna and Saikat Sinha regarding computational techniques. This work was supported by Simons Foundation Grant No.\ 376198.

\end{document}